\documentclass[usenatbib]{mnras}
\usepackage{graphicx}
\usepackage{float}
\usepackage{hyperref}
\usepackage{color}

\newcommand{\be}[1]{\begin{equation} \label{eq:#1}}
\newcommand{\ee}{\end{equation}}
\newcommand{\ba}[1]{\begin{eqnarray} \label{eq:#1}}
\newcommand{\ea}{\end{eqnarray}}

\newcommand{\ra}{\ifmmode{
\mathbf{ \hat{r}_a}
} \else
$\mathbf{ \hat{r}_a}$\fi}

\newcommand{\rb}{\ifmmode{
\mathbf{ \hat{r}_b}
} \else
$\mathbf{ \hat{r}_b}$\fi}

\newcommand{\rc}{\ifmmode{
\mathbf{ \hat{r}_c}
} \else
$\mathbf{ \hat{r}_c}$\fi}

\newcommand{\B}{\ifmmode
\mathbf{B}\else {\bf B}\fi}

\newcommand{\bobs}[2]{\ifmmode{
\mathbf{b}^{#1}_{#2}
} 
\else
\bf{b}$^{#1}_{#2}$\fi
}

\newcommand{\solrad}{\ifmmode{R}_{\rm S}\else${R}_{\rm S}$\fi}
\newcommand{\solmas}{\ifmmode{M}_{\rm S}\else${M}_{\rm S}$\fi}

\newcommand{\tintu}{\ifmmode{\rm erg~cm^{-2}~s^{-1}sr^{-1}}\else 
  erg~cm$^{-2}$~s$^{-1}$~sr$^{-1}$\fi}
\newcommand{\fluxu}{\ifmmode{\rm erg~cm^{-2}~s^{-1}}\else 
  erg~cm$^{-2}$~s$^{-1}$\fi}
\newcommand{\velu}{$\,$km$\,$s$^{-1}$}

\newcommand{\wave}{\ifmmode{\lambda} \else$\lambda$\fi}

\newcommand\lta { \mathrel {\hbox to 0pt {\lower 3.7pt \hbox{$\sim$}
      \hss} \raise 1.7pt \hbox{$<$}}}
\newcommand\gta { \mathrel {\hbox to 0pt {\lower 3.7pt \hbox{$\sim$}
      \hss} \raise 1.7pt \hbox{$>$}}}

\newcommand{\intu}{\ifmmode{\rm erg~cm^{-2}~s^{-1}sr^{-1} \AA^{-1}}\else 
 erg~cm$^{-2}$~s$^{-1}$~sr$^{-1}$
 ~\AA$^{-1}$\fi}

\newcommand{\intunu}{\ifmmode{\rm erg~cm^{-2}~s^{-1}sr^{-1} \AA^{-1}}\else 
 erg~cm$^{-2}$~s$^{-1}$~sr$^{-1}$
 ~Hz$^{-1}$\fi}

\newcommand{\figstart}{
\begin{figure*}
\includegraphics[width=\linewidth]{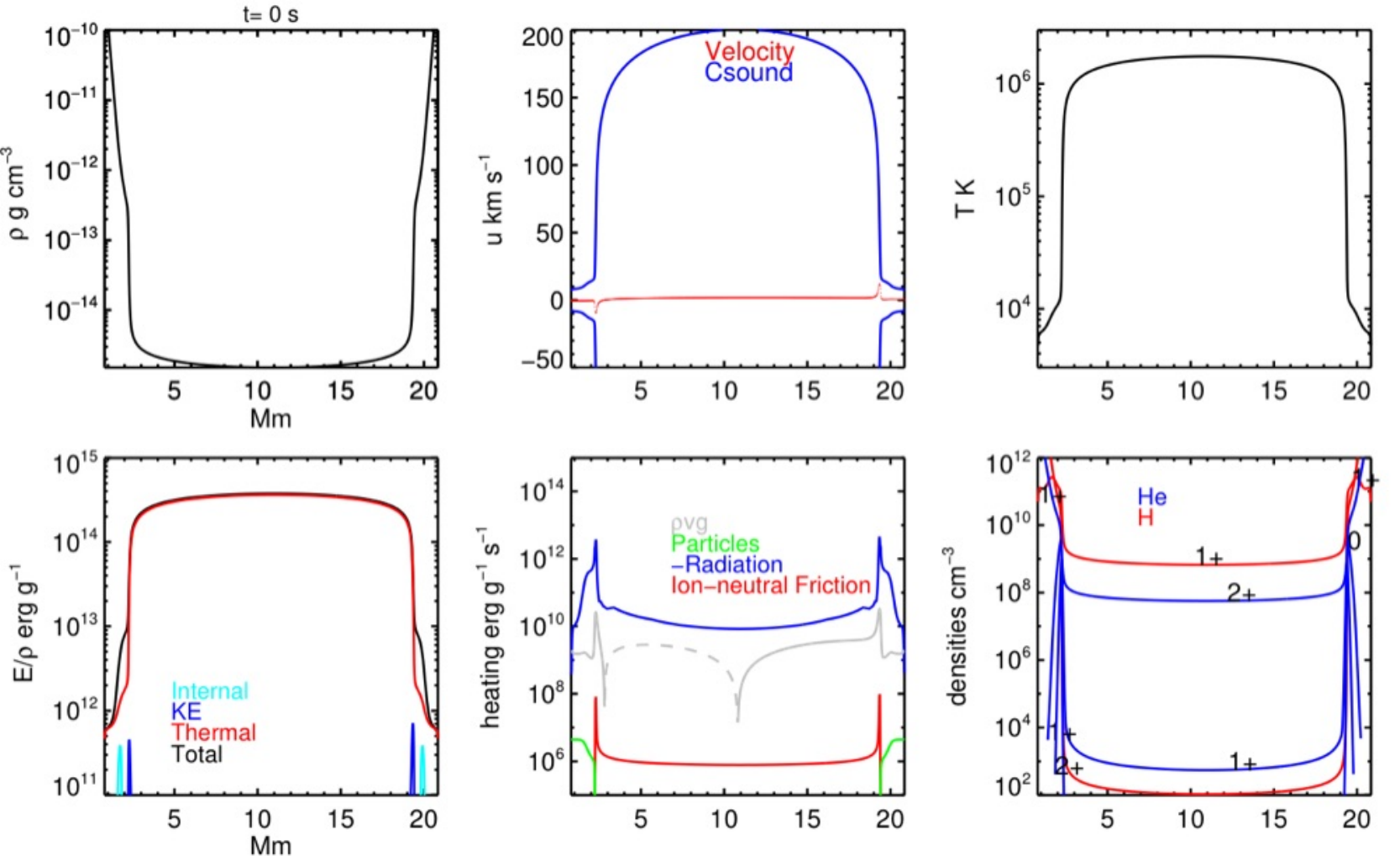}  
\caption{Initial state
for calculations with $L=20$ Mm.   The curves labeled particles"  in the bottom middle panel is the heating of neutrals used at the footpoints to support the modeled ``chromosphere'' at the base. The upper middle panel shows flow speeds spanned by 
sound speeds in the forward and backward (negative values)  directions.  Note that 
the energy densities of the intermediate waves are not
plotted.  Negative values of are shown as dotted lines in these logarithmically-scaled plots. } 
\label{fig:start} 
\end{figure*}
}

\newcommand{\figec}{
\begin{figure}
\includegraphics[width=\linewidth]{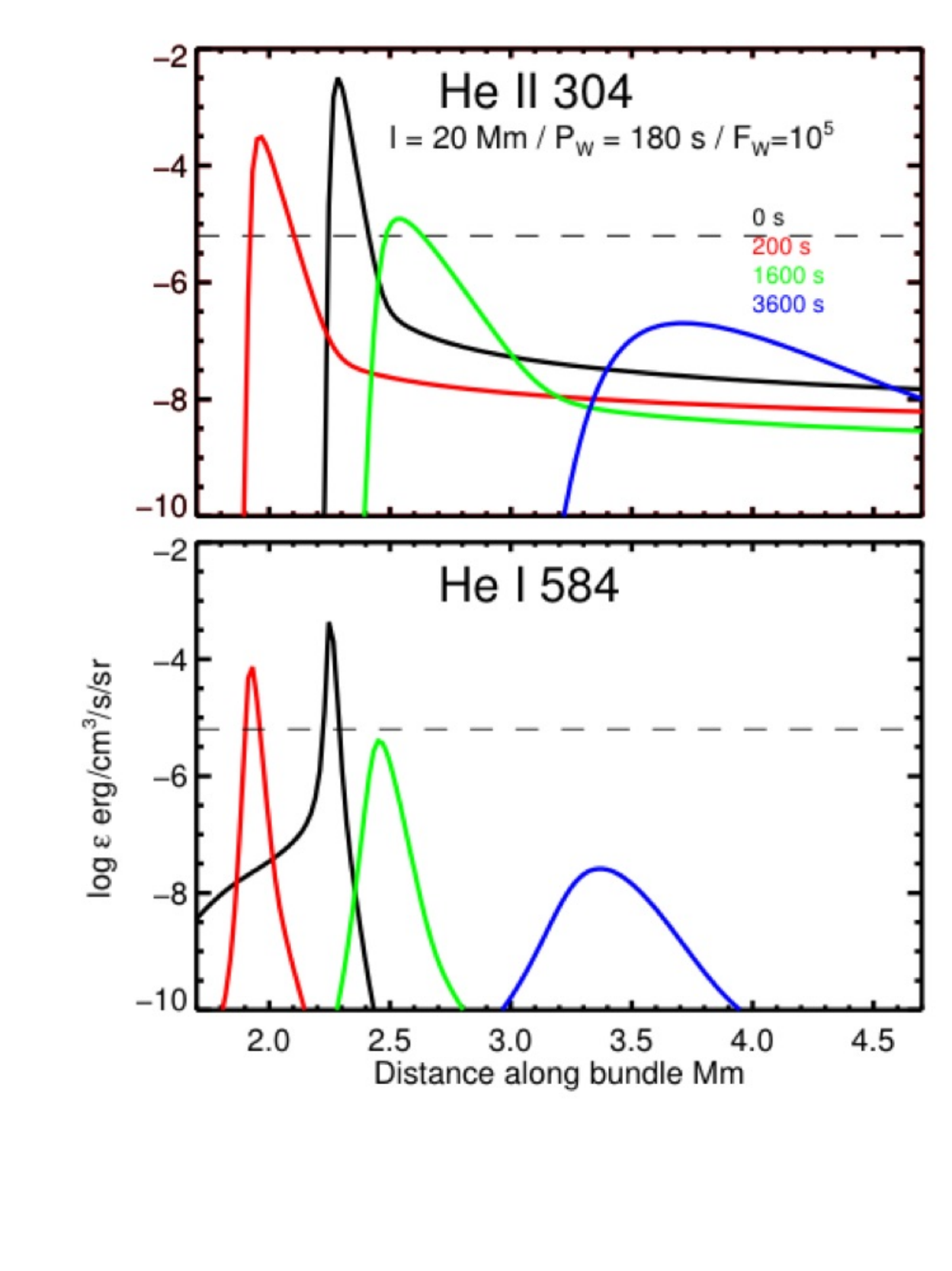}  
\caption{Emission coefficients for resonance lines of He and He$+$ are shown as a function of
position along the bundle, at four different times.
The horizontal dashed line
indicates the emission coefficient needed to 
contribute 10\%{}
of observed 
intensities, assuming a 
bundle thickness
(cross-bundle path length for integration, equation~\ref{eq:intense}) of 100 km.
} \label{fig:ec} 
\end{figure}
}

\newcommand{\figinter}{
\begin{figure*}
\includegraphics[width=\linewidth]{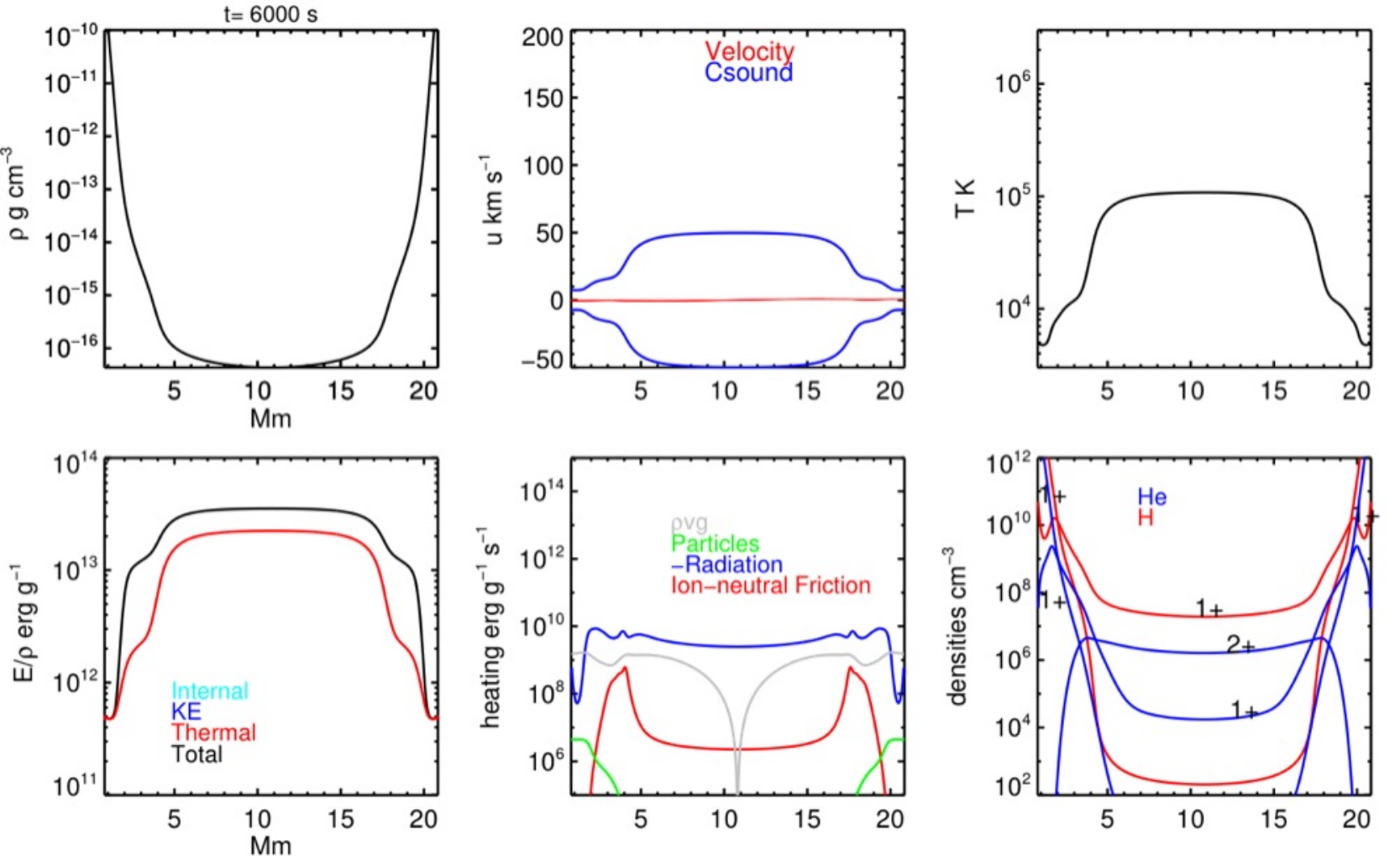}
\caption{
State of calculation 20 after t= 100 minutes.  Notice that the ratio of neutral hydrogen to protons has increased to  $\approx10^{-5}$ 
following advection and recombination.
This leads to the significant heating rate shown in red in the middle-lower panel of between  $10^8$ and $10^{10}$ erg~g~s$^{-1}$, between 4 and 17 Mm along the bundle. 
In contrast, the initial state (Figure~\protect\ref{fig:start}) has a ratio of neutral to ionized hydrogen of $10^6$ with no significant ion-neutral heating.  
} 
\label{fig:inter} 
\end{figure*}
}

\newcommand{\figib}{
\begin{figure}
\includegraphics[width=\linewidth]{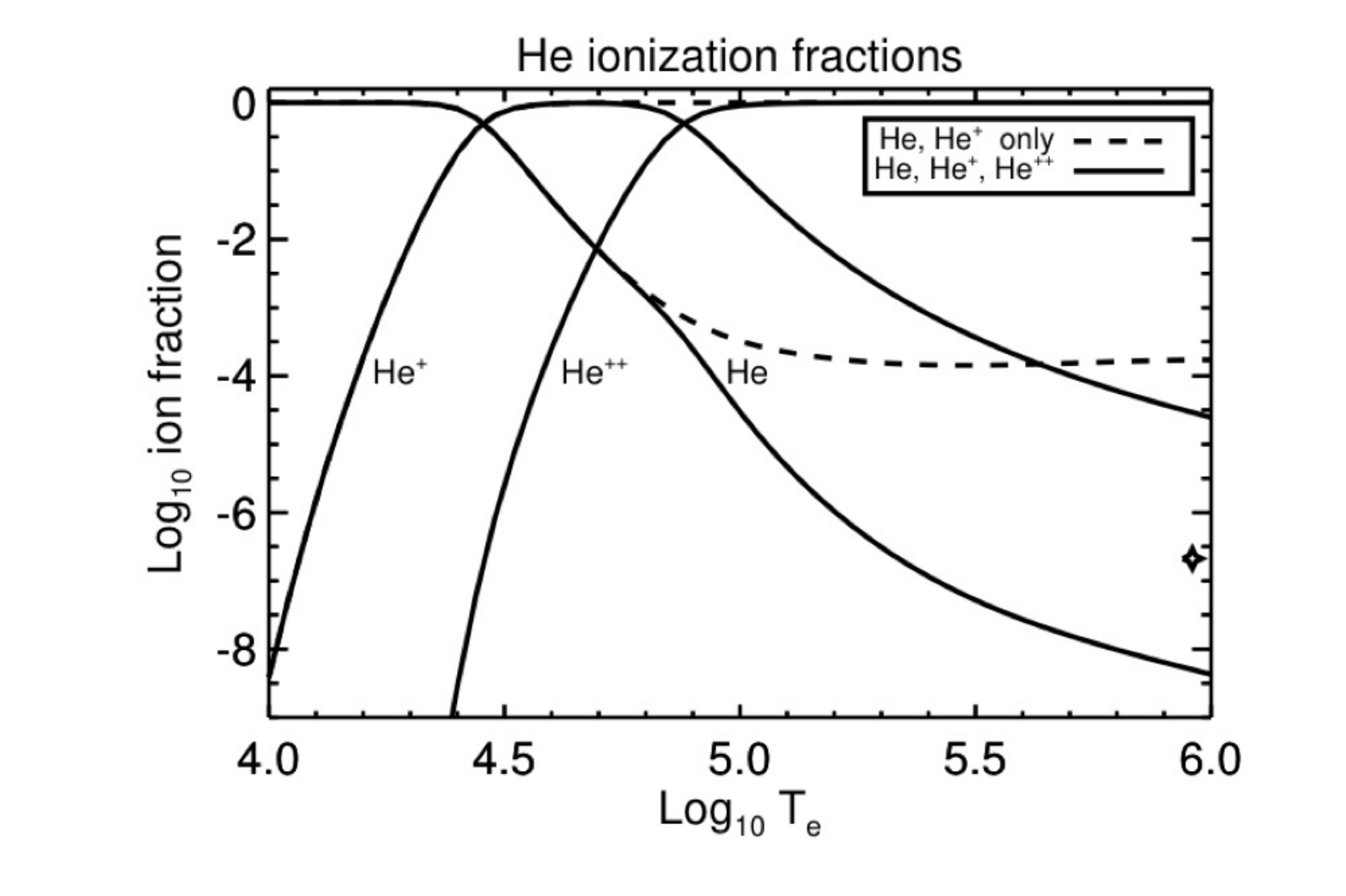}  
\caption{Ionization equilibrium 
calculations for helium, using 
only He, He$^+$ (dashed lines) and using all
three stages of ionization (solid).  Only collisions with electrons have been included. 
The difference between the dashed and solid lines for neutral helium reveals the erroneously high He neutral populations
computed by \citet{2005ARep...49.1009Z}. The star symbol shows the ion fraction for neutral hydrogen near 10$^6$K.
} \label{fig:ib}
\end{figure}
}

\newcommand{\figfinal}{
\begin{figure*}
\includegraphics[width=\linewidth]{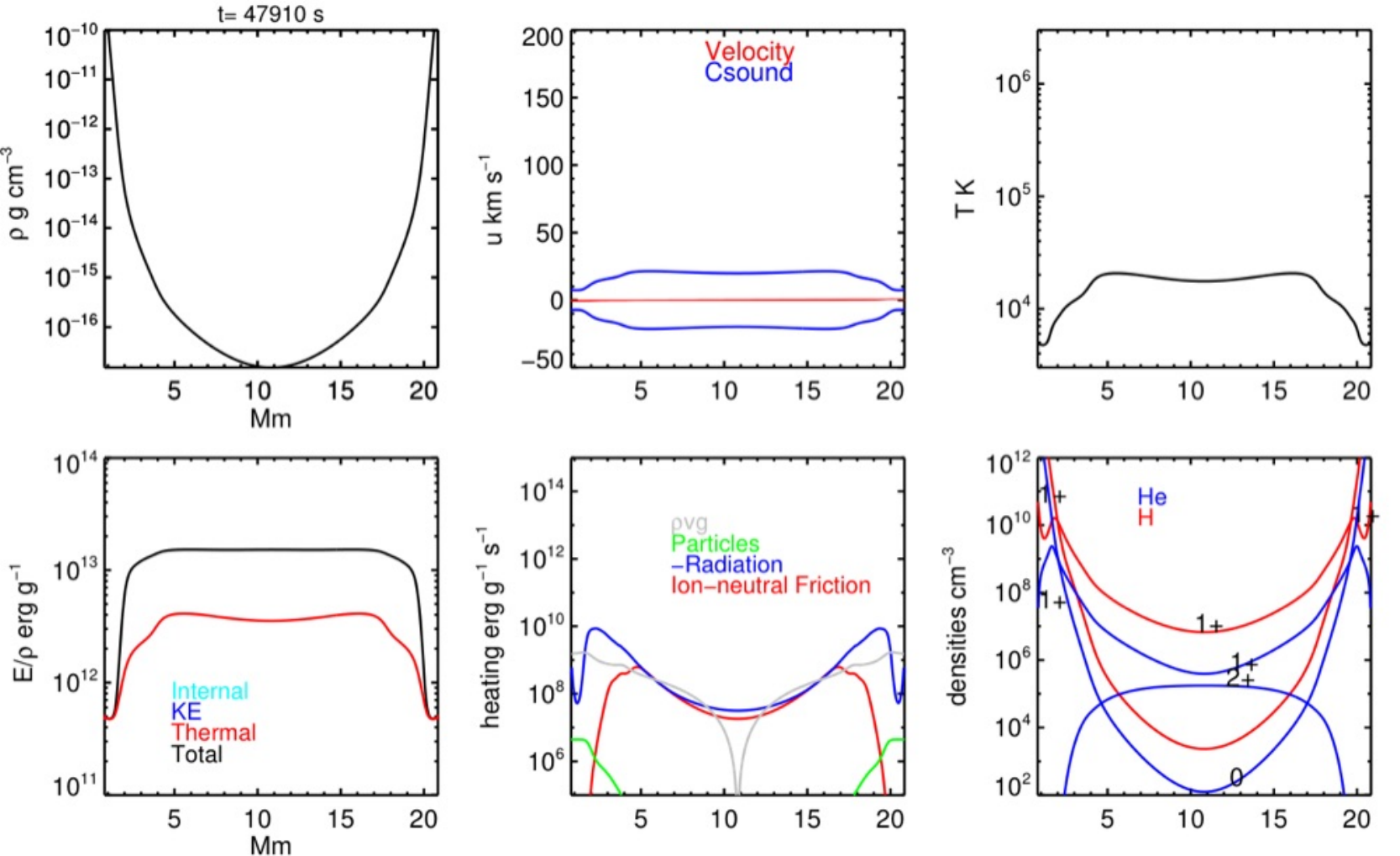}  
\caption{
State of calculation 20 after t=800 minutes.} \label{fig:final}
\end{figure*}
}

\newcommand{\figcatas}{
\begin{figure*}
\includegraphics[width=\linewidth]{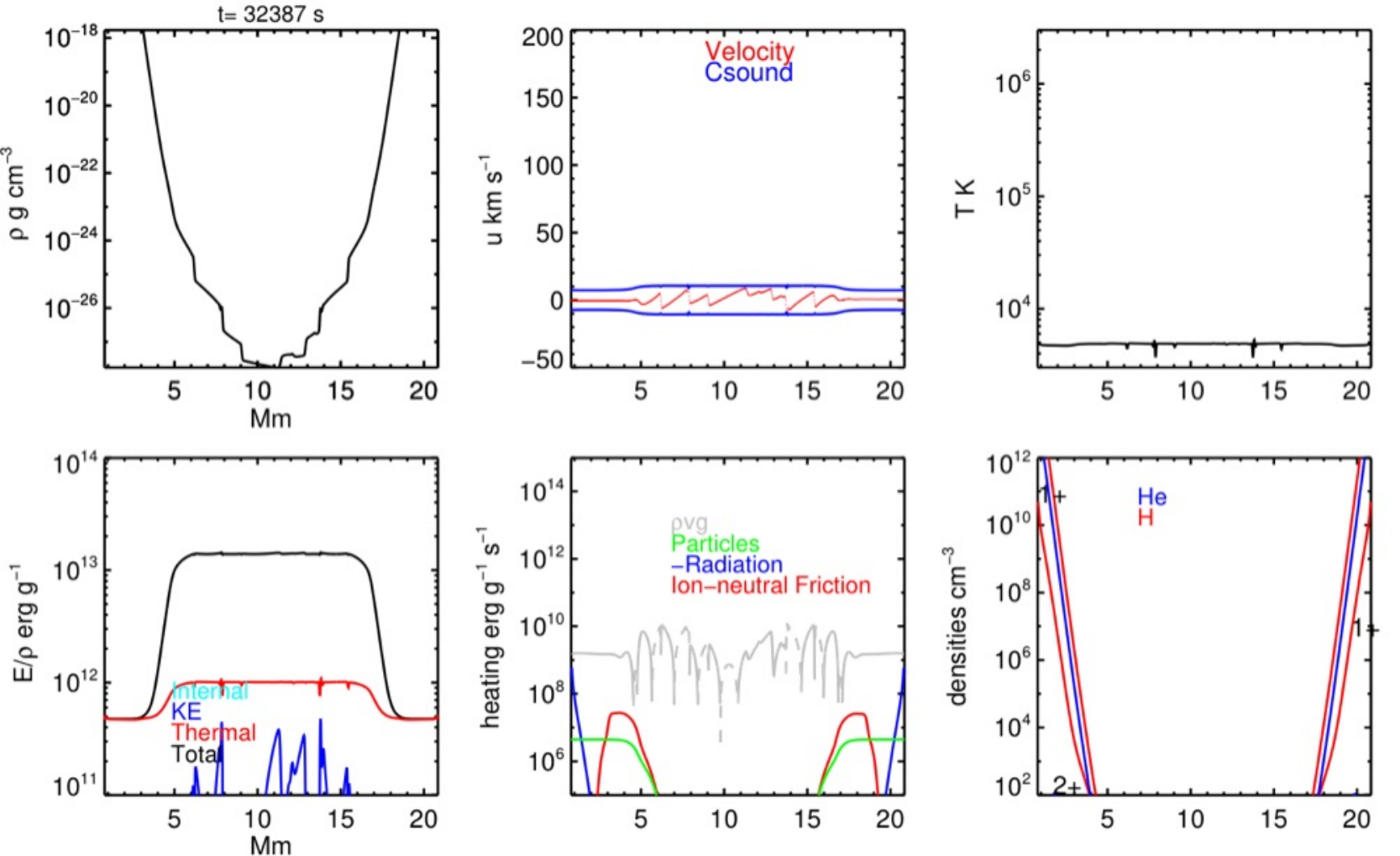}  
\caption{
State of calculation 13 after t=396 minutes.   The fluid calculation has collapsed to
a state where mean free paths
exceed the size of the system
because of the catastrophic loss of support for the atmosphere by insufficient 
ion-neutral heating.  \new{This implies
that a ``collisionless'' approximation to the transport 
problem is more appropriate (see text)}. 
} \label{fig:catas}
\end{figure*}
}

\newcommand{\fighalfg}{
\begin{figure*}
\includegraphics[width=\linewidth]{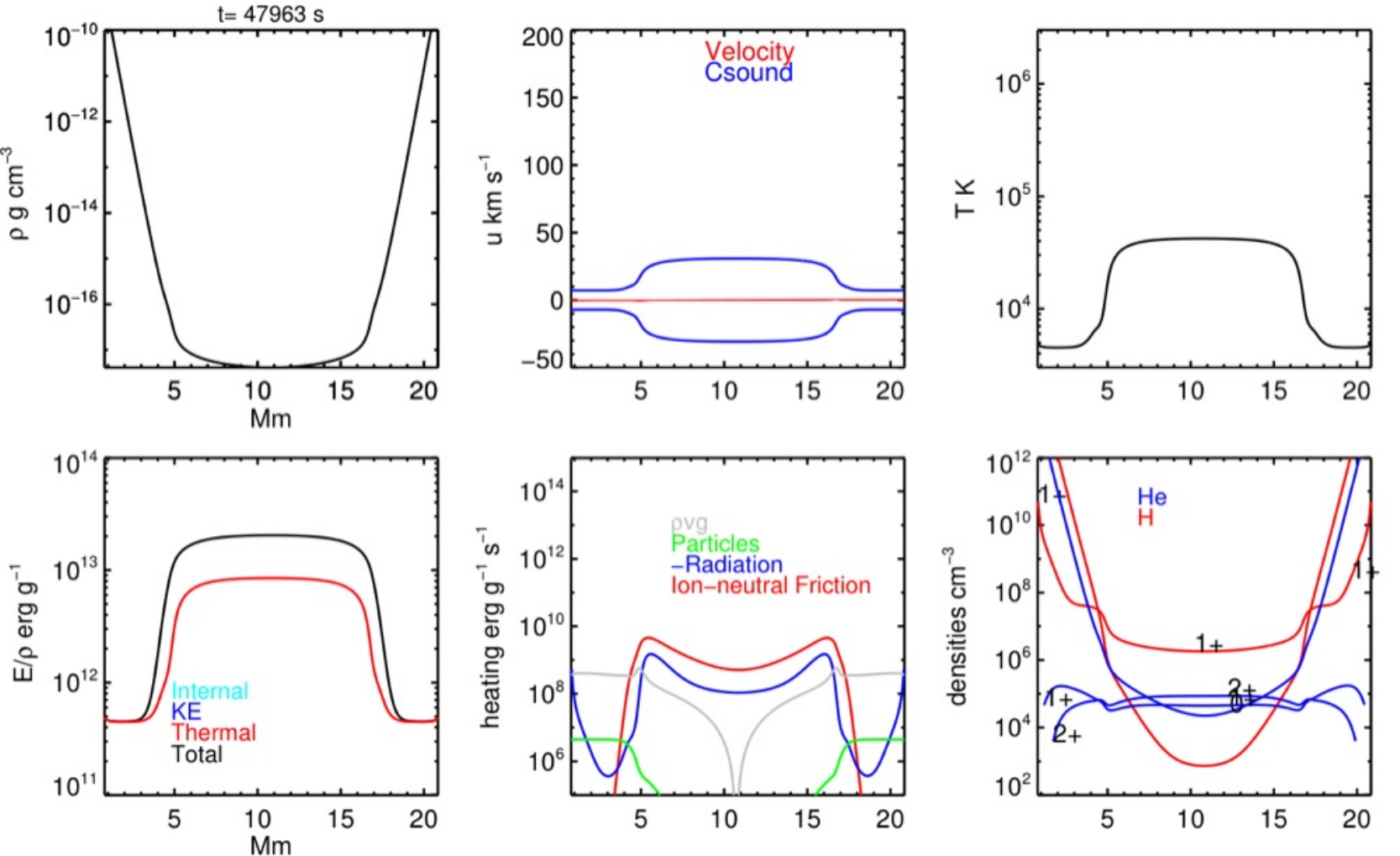}  
\caption{
State of calculation 31 after t= 800 minutes, a calculation identitical to
the collapsed case (Figure~\ref{fig:catas})
but using a gravity half that of the Sun in the radial direction, 
for a bundle lying in a plane inclined at
60$^\circ$ to the vertical. } \label{fig:halfg}
\end{figure*}
}

\newcommand{\figprof}{
\begin{figure}
\includegraphics[width=\linewidth]{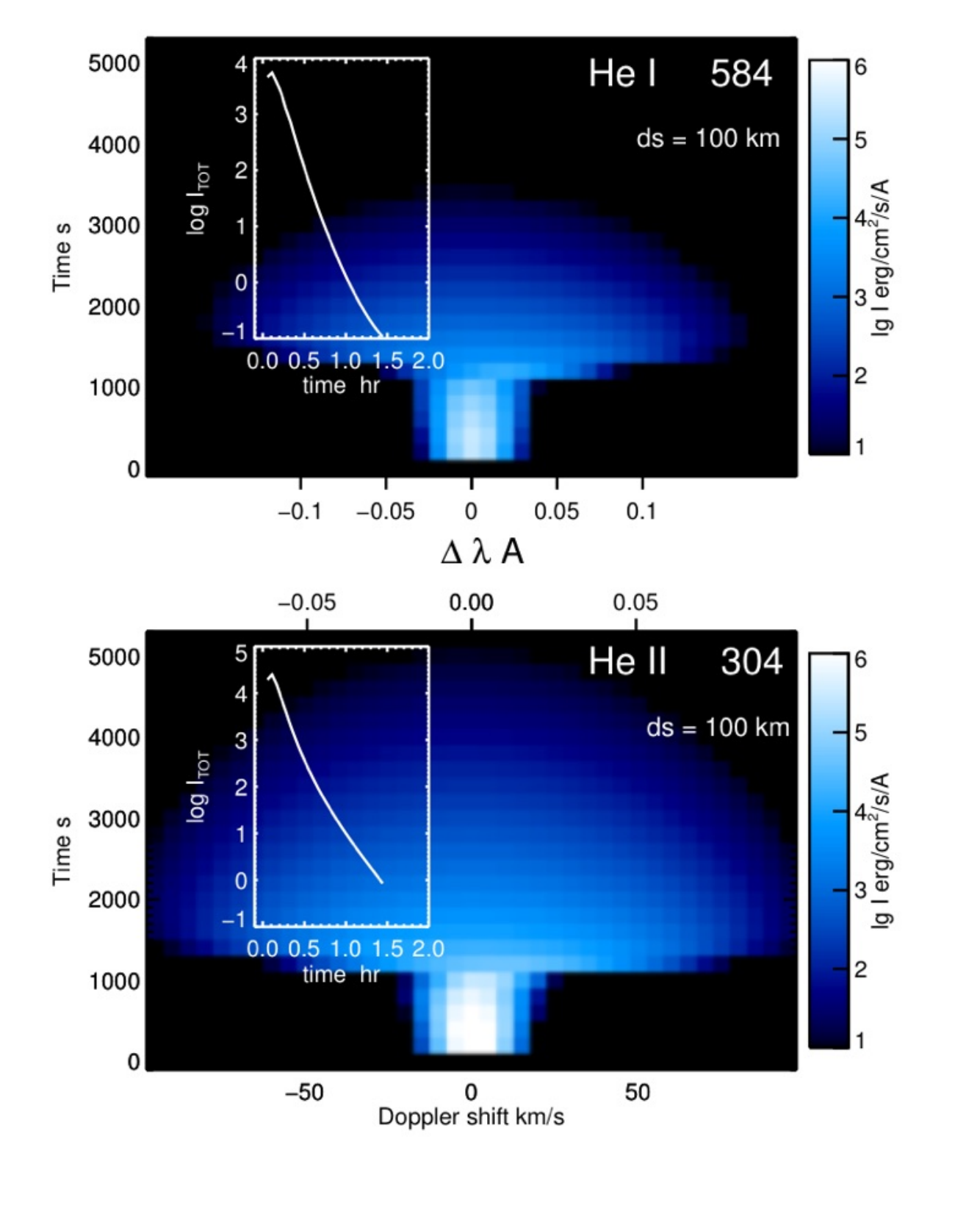}  
\caption{
Evolution of the He and He$^+$ resonance line profiles 
and integrated intensities (inset plots).  Units
are \intu{} (images) and 
\tintu{} (line plots).  The profiles shown were computed 
by identifying the brightest features along the bundle,
and stacking these up to form a plot in time.  This
is what an instrument with a modest angular resolution 
of $\ge1$ Mm would observe.  
Data are shown for the footpoint near $x=0$
from 
calculation 22 of Table~\ref{tab:calcs}.
The path length for integration (equation~\ref{eq:intense}) was 100 km.
} 
\label{fig:prof}
\end{figure}
}

\newcommand{\figeq}{
\begin{figure}
\includegraphics[width=\linewidth]{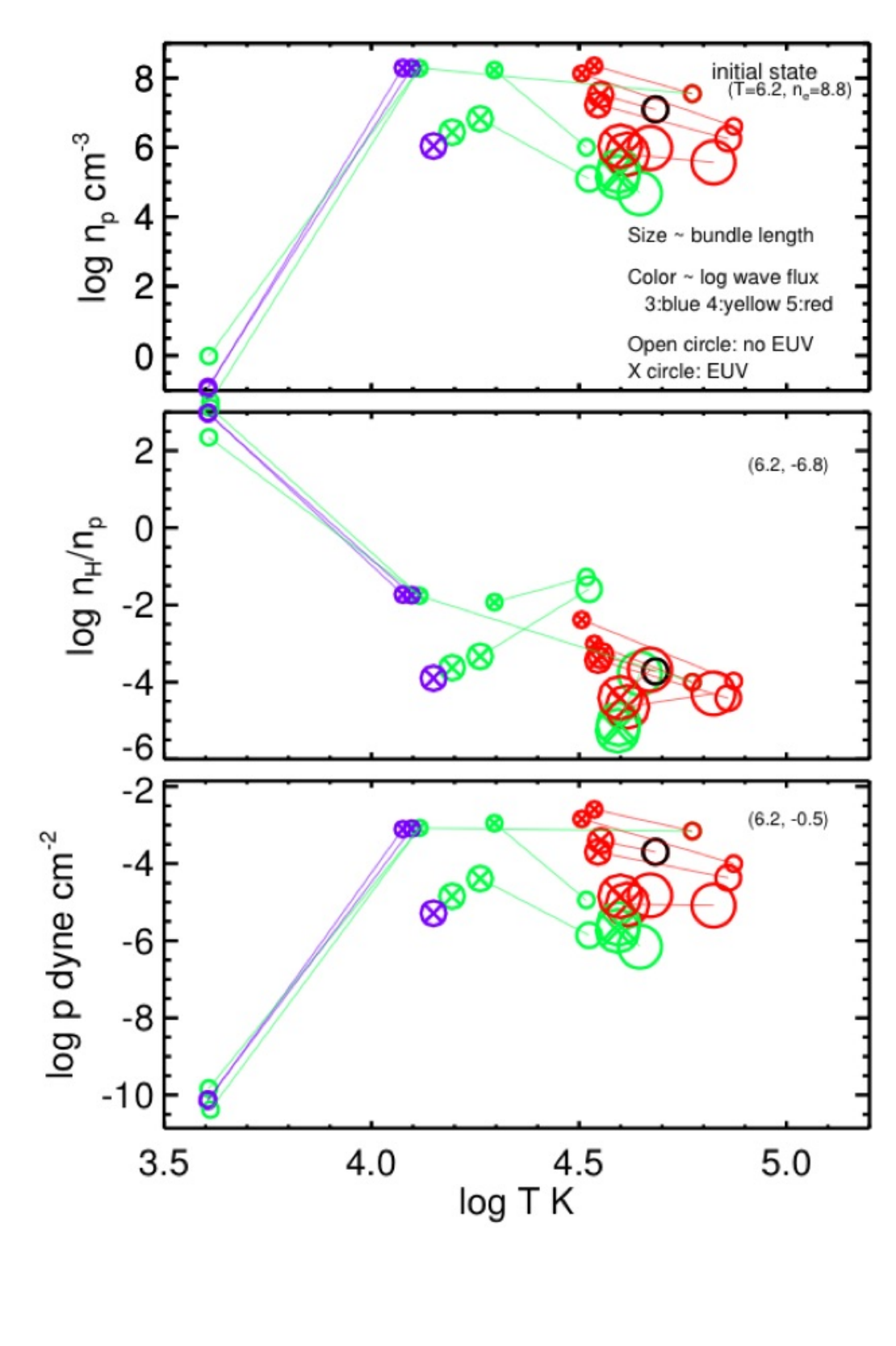}
\caption{Scatter plots are shown using the numbers in
Table~\ref{tab:calcs}, only for those calculations marked ``EQ''. The size of circles indicates the length $L$ of the bundle for each calculation.  The colors
represent the amplitude $\xi$, blue =0.6, yellow=
3, red=10 \velu{}.
The values 
are for the geometric mid-point (i.e. highest point) of each bundle. 
The lines plotted connect points where  calculations were made with and then without EUV photoionization and
heating. 
Calculation 20 
is marked in black.
}\label{fig:eq}   
\end{figure}
}

\newcommand{\tabcalcs}
           { \protect 
  \begin{table*}
      \caption{List of  modeled configurations and their outcomes. }
      \begin{tabular}{lllllrrl}
      \hline
      Calc. & 
  $L$ &$P_W$ &  log$_{10} F_W$ &         log$_{10}T$ & log$_{10}n_p$ &log$_{10}\frac{nH}{np}$ & outcome\\
 & Mm &  seconds & erg~cm$^{-2}$s$^{-1}$  & K & cm$^{-3}$ & \\
  \hline
  initial& 20 &  \ldots    &    \ldots     &         6.24&      8.82   &  -6.80  &  \\ %,20cor
   1 &   10 & 1000 &    3 &            3.61&     -0.97&      2.99&  EQ\\%1010003
   2 &   10 & 1000 &    3+EUV &       4.08&      8.29&     -1.73&  EQ\\%1010003e
   3 &   10 & 1000 &    4      &       3.61&     -0.01&      2.35&  EQ\\%1010004
   4 &   10 & 1000 &    4+EUV &       4.12&      8.28&     -1.76&  EQ\\%1010004e
   5 &   10 & 1000 &    5      &       4.87&      6.61&     -3.97&  EQ\\%1010005
   6 &   10 & 1000 &    5+EUV &       4.51&      8.13&     -2.38&  EQ\\%1010005e
   7 &   10 &  180 &    3 &            3.61&     -0.91&      2.96&  EQ\\%101803
   8 &   10 &  180 &    3+EUV &       4.10&      8.28&     -1.74&  EQ\\%101803e
   9 &   10 &  180 &    4 &            4.52&      6.00&     -1.28&  EQ\\%101804
  10 &   10 &  180 &    4+EUV &       4.30&      8.23&     -1.93&  EQ\\%101804e
  11 &   10 &  180 &    5 &            4.77&      7.55&     -4.00&  EQ\\%101805
  12 &   10 &  180 &    5+EUV &       4.54&      8.35&     -3.01&  EQ\\%101805e
  13 &   20 & 1000 &    4 &            3.69&          &          &      23800\\ %,2010004
  14 &   20 & 1000 &    4+EUV &       4.19&      6.44&     -3.63&  EQ\\%2010004e
  15 &   20 & 1000 &    5 &            4.86&      6.25&     -4.41&  EQ\\%2010005
  16 &   20 & 1000 &    5+EUV &       4.55&      7.23&     -3.43&  EQ\\%2010005e
  17 &   20 &  180 &    3 &            3.67&          &          &     20200\\ %,201803
  18 &   20 &  180 &    3+EUV &       4.15&      6.04&     -3.90&  EQ\\%201803e
  19 &   20 &  180 &    4 &            4.52&      5.09&     -1.59&  EQ\\%201804
  20 &   20 &  180 &    4+EUV &       4.26&      6.82&     -3.33&  EQ\\%201804e
  21 &   20 &  180 &    5 &            4.68&      7.09&     -3.72&  EQ\\%201805
  22 &   20 &  180 &    5+EUV &       4.55&      7.52&     -3.31&  EQ\\%201805e
  23 &   40 & 1000 &    4 &            3.94&          &          &      36400\\ %,4010004
  24 &   40 & 1000 &    4+EUV &       4.59&      5.14&     -5.26&  EQ\\%4010004e
  25 &   40 & 1000 &    5 &            4.82&      5.57&     -4.30&  EQ\\%4010005
  26 &   40 & 1000 &    5+EUV &       4.62&      5.80&     -4.64&  EQ\\%4010005e
  27 &   40 &  180 &    4 &            4.65&      4.68&     -3.78&  EQ\\%401804
  28 &   40 &  180 &    4+EUV &       4.60&      5.31&     -5.09&  EQ\\%401804e
  29 &   40 &  180 &    5 &            4.67&      5.98&     -3.68&  EQ\\%401805
  30 &   40 &  180 &    5+EUV &       4.60&      6.03&     -4.41&  EQ\\%401805e
  31 &   20 & 1000 &    4   g/2 &       4.62&      6.26&     -3.13&  EQ\\%2010004g
   32 &   10 & 180 &    4 c &            3.61&     -1.33&      3.12&  EQ\\%1010004c4
   33&   10 & 180 &    5 c &          4.77&      7.55&     -4.00&  EQ\\%1010004c
    		          \hline
		          \end{tabular} \break
		         The calculations listed all begin with 
a near-equilibrium coronal model (the
model for $L=20$ Mm is listed first).    ``+EUV"
indicates that EUV photo-ionization of H and He was included.  Calculation 31 is identical to 13 except that half of the 
vertical solar gravity was used, assuming
the bundle lies in a plane inclined at $60^\circ$ to the local vertical.  Calculations 32 and 33 were initiated from
the final (very cool) solution of calculation 3.  \new{The numbers in the last column refer to the 
time at which the calculation 
reached a collapsed state from which further calculations were
not possible.} 
\label{tab:calcs} 
  \end{table*}
}

\newcommand{\tabflux}
           { \protect 
  \begin{table}
          \caption{Adopted intermediate wave flux densities }
      \begin{tabular}{lll}
      \hline
$F_W$ & Initial $\xi$ & B\\
\fluxu{} & \velu{}& G\\
\hline
$10^3$  & 1 & 10\\
$10^4$  & 3 & 10\\
$10^5$  & 10& 10\\
\hline\\
\end{tabular}
\label{tab:fluxes}
  \end{table}
}

\newcommand\new[1]{{\bf #1}}

\catcode`_=\active
\newcommand_[1]{\ensuremath{\sb{\mathrm{#1}}}}
\catcode`^=\active
\newcommand^[1]{\ensuremath{\sp{\mathrm{#1}}}}

\title{Inevitable consequences of ion-neutral damping of intermediate MHD waves in Sun-like stars}
\author[P.G. Judge]{Philip G. Judge\thanks{E-mail: judge@ucar.edu}\\ %, and J.R. Kuhn\\
High Altitude Observatory,\\
National Center for Atmospheric Research,\\
Boulder CO 80307-3000,
 USA } 

\date{Accepted . Received ; in original form }

\pubyear{2020}

\begin{document}
\label{firstpage}
\pagerange{\pageref{firstpage}--14}

\maketitle

%###############################################################################
%
%     ABSTRACT
%
%###############################################################################

\begin{abstract}
In the context of the solar atmosphere, we re-examine 
the role of 
of neutral and ionized 
 species in dissipating the  ordered
 energy of intermediate-mode MHD waves into heat.
 We solve conservation equations for the hydrodynamics and for hydrogen and helium ionization stages, along closed tubes of magnetic field.
 First, we examine the evolution of coronal plasma under conditions where coronal heating has abruptly ceased.   
 We find that cool ($< 10^5$K) 
 structures are formed lasting for several hours. MHD waves of modest amplitude can heat the plasma through ion-neutral collisions with sufficient energy rates to support the plasma against gravity.
  Then we examine a calculation  starting from 
  a cooler atmosphere.  The calculation shows that warm ($> 10^4 $) K long ($>$ several Mm)  tubes of plasma  arise  by
  the same mechanism.
    We speculate on the relevance of these
    solutions to observed properties of the Sun and similar stars whose atmospheres are permeated with emerging  magnetic fields and stirred by convection. Perhaps this elementary process might help explain the presence of  ``cool loops'' in the solar transition region and the production of broad components of transition region lines.
 The  production of ionized hydrogen from such a simple and perhaps inevitable mechanism may be an important step towards finding the more complex mechanisms needed to generate coronae with temperatures in excess of 10$^6$K, independent of a star's metallicity.   
\end{abstract}


\section{Statement of the problem}
\label{sec:statement}

This paper re-examines influences of intermediate MHD waves on the
solar atmosphere above the stratified chromosphere.
Direct observational signatures of such waves with periods of several
minutes has been documented for more than a decade
\citep{Tomczyk+others2007, 2011Natur.475..477M}.   When account is taken
of likely wave amplitudes inferred from
measured widths of chromospheric and coronal emission lines
at visible 
\citep{Billings1966} and UV wavelengths \citep{Cheng+Doschek+Feldman1979,
    2000ApJ...529..599D,2016ApJ...820...63B},
the energy flux of such waves
appears ample to support radiative and other losses from
the solar corona.   
It is well-established that frictional and viscous forces are
unable to dissipate these waves quickly enough in
highly ionized plasma, unless dynamical processes,
such as phase-mixing and resonant absorption,
can generate structure on scales far below
the typical wavelengths. Some difficulties
with such proposed mechanisms have been highlighted by, for example, 
\citet{Parker1991,Ofman+Klimchuk+Davila1998,2016ApJ...823...31C}, and
the complexity of the problems that these mechanisms entail
make this an active research area today.

However, two articles have re-emphasized the role
of ion-neutral friction in heating the corona
\citep{2005ARep...49.1009Z,2020A&A...635A..28W}.
In both studies, the fractions of neutral atoms
greatly exceed those expected under conditions of
coronal ionization equilibrium.  Figure~\ref{fig:ib} 
shows ionization fractions computed under optically thin,
radiation-free conditions using the {\tt HAOS-DIPER} package
\citep{Judge2007a}.  
In 
\citet{2005ARep...49.1009Z},  the ionization fraction
of neutral helium above 10$^6$K exceeds $10^{-5}$, where
our calculations indicate values below $10^{-8}$.  The
difference is seen explicitly through their equations
(23) and (24), which neglects ionizations from
He$^+$ to He$^{2+}$.  Figure~\ref{fig:ib} highlights
how this affects the coronal fractions of neutral
He, the dashed line is equivalent to the calculation
of \citet{2005ARep...49.1009Z}.

\citet{2020A&A...635A..28W}
made 2D MHD calculations in a two fluid model of the solar atmosphere
including ion-neutral collisions. Their calculations show
even higher neutral fractions in the corona than were found by
\citet{2005ARep...49.1009Z}, because no 
  sources or sinks of ionization and recombination by collisions with
  electrons were treated. Instead the authors including only the advection term in their
  solutions for number densities.

\figib

One dimensional (i.e. field-aligned)
hydrodynamical models of the chromosphere and corona have been studied
for several decades. Early  work includes 
\citet{1982ApJ...254..349O,
  1984MmSAI..55..749P,1987ApJ...319..465M,
  1988A&A...200..153K}, and more recent
advances including adaptive numerical resolution, first
applied by \citet{1983ApJ...265..483M} to the transition region,
are summarized by, e.g.,
\citet{Hansteen1993, 2013ApJ...770...12B, 2018ApJ...856..178P}.  Two-
 and three- dimensional models
 \citep{2002A&A...382..639R,2011A&A...531A.154G,
   2017ApJ...834...10R} including magneto-hydrodynamics, radiative transfer, ion-neutral
physics  and field-aligned connections to the solar corona have been
developed. The latest 3D models required enormous efforts by many people over
years. They currently represent the closest numerical work to genuine
``simulations'' of the solar atmosphere (as opposed to numerical
experiments), and represent the state of the art.

\figstart

Yet, we believe that a class of quasi-stable solutions may have been missed.
\citet{1993ApJ...412..375L} concluded from elementary 
considerations that coronal heating  ``must be impulsive in
onset.''  Many papers have provided observational and
theoretical evidence for coronal heating that is
intermittent in space and time \citep[see reviews by][]{Klimchuk2006SoPh,
2012RSPTA.370.3217P,2015RSPTA.37340269D}.
This result is a consequence of the special
  conditions needed to generate 
the necessarily small scales, which 
arise naturally under  local and non-local conditions
from intermediate wave resonances and the slower build up of magnetic stresses 
\citep{Ionson1978,Parker1988}.   Once
this heating has occurred, the small structures will  
evolve rapidly, which will either dampen or enhance the small-scale
energy
dissipation. Thus, if the development of small scales is indeed the
pre-requisite for coronal heating, then the heating process inevitably
must be intrinsically \textit{dynamic and intermittent}.

Our goal therefore is to examine the dynamics of plasma in
  tubes of magnetic flux  \textit{between the episodes of strong coronal
  heating}, such as when the resonance condition of Ionson or the
``critical angle'' condition for Parker's nano-flares are not met.  We
build a model containing the fewest ingredients that can transport and
deposit energy above the chromosphere.
All stars that rotate and convect will have varying
emergent magnetic fields \citep{Spruit2011}, and all Sun-like stars in
the solar neighborhood of the Galaxy possess hot coronae
\citep{Schmitt1997}, and an abundance of ionized
atoms.  Between episodes of coronal heating, all such
stars must experience the effects of ion-neutral collisions
on intermediate waves.  Thus we re-examine
intermediate MHD waves dampened by ion-neutral
collisions.

%%%%%%%%%%%%%%%%%%%%%%%%%%%%%%%%%%%%%%%%%%%%%%%%%%

\begin{table}
  \caption{Logarithmic values of typical physical parameters }
  \begin{tabular}{llrrr}
    \hline
    \hline
    Quantity & unit &  \multicolumn{3}{c}{values} \\
    \hline
    $T$ &K &    4.0 &    5.0 &    6.0  \\
    $n_e$ &cm$^{-3}$ &   10.4 &    9.4 &    8.4  \\
    $\rho$ &g cm$^{-3}$ &  -13.0 &  -14.0 &  -15.0  \\
    B & G   &    \multicolumn{3}{c}{   \hrulefill\   1.0   \hrulefill\   } \\
    $\Omega_i$ &rad s$^{-1}$ &    \multicolumn{3}{c}{   \hrulefill\   5.0   \hrulefill\   }  \\
    $2\pi/\Omega_i$ &s &   \multicolumn{3}{c}{   \hrulefill\   -4.2   \hrulefill\   }  \\
    \\
    $\tau_{ee}$ & s   &   -6.0 &   -3.5 &   -1.0  \\
    $\tau_{ii}$ & s   &   -4.4 &   -1.9 &    0.6  \\
    $\tau_{ie}$ & s   &   -2.5 &   -0.5 &    1.5  \\
    $\tau_{in}^*$ & s  &   -2.9 &    2.5 &    4.6  \\
    $\tau_{in}$ & s   &   -2.4 &    3.0 &    5.1  \\
\\
    $\tau_{rec}^a$ & s  &    2.0 &    3.5 &    5.0  \\
    $\tau_{ion}^b$ & s  &    4.3 &   -1.3 &   -1.4  \\
    $\tau_{pion}^c$ & s &    \multicolumn{3}{c}{   \hrulefill\   2.0   \hrulefill\   } \\
    $\tau_{tion}^d$ & s  &    2.0 &   -1.3 &   -1.4  \\
    $n_H/(n_H+n_p)$ &  &   -0.3 &   -4.8 &   -6.5  \\
    \\
    $L^e$ & cm  &    \multicolumn{3}{c}{   \hrulefill\   9.7   \hrulefill\   } \\
    $F_R^f$& erg cm$^{-2}$ s$^{-1}$ &    \multicolumn{3}{c}{   \hrulefill\   5.4   \hrulefill\   }  \\
    $F_R/L^g$ & erg cm$^{-3}$ s$^{-1}$ &   \multicolumn{3}{c}{   \hrulefill\   -4.3   \hrulefill\   } \\
    $\tau_{cool}^h$ & s &    \multicolumn{3}{c}{   \hrulefill\   3.3   \hrulefill\   } \\
    $L/c_S$  & s &    3.7 &    3.2 &    2.7  \\
    $L/c_A$  & s &    2.7 &    2.2 &    1.7  \\
    Wave period P$_W^i$ &s &    \multicolumn{3}{c}{   \hrulefill\   2.3,    3.0   \hrulefill\   } \\
    Wave freq. $\omega$ & rad s$^{-1}$ &    \multicolumn{3}{c}{   \hrulefill\   -1.5, -2.2   \hrulefill\   } \\
    \hline
  \end{tabular}
  \break
  Physical properties of the calculations
  are listed as base 10 logarithms.  All quantities refer to
  a pure hydrogen atmosphere at a constant pressure of
  $p=0.07$ dyne~cm$^{-2}$. The quantities $\tau_{in}$ refer to
  time scales for collisions on ion $i$ by neutral $n$, and so
  forth.  $^*$The value is for momentum-changing charge-transfer transitions
  \citep[eq. A17 of][]{Gilbert+Hansteen+Holzer2002}, the unsuperscripted
  value of $\tau_{in}$ is for elastic collisions \citep{1977RSPSA.353..575H}. 
$^a$Recombination times for protons colliding with electrons,
  $^b$ionization times for hydrogen atoms colliding with electrons,
  both estimated using data from \citet{Allen1973}.
  $^c$photionization times,
 from \citet{1995JGR...100.3455O} (for helium, we use
\citealp{2014arXiv1411.4826S}). $^d$total ionization times.  $^e$Loop length, $^f$energy flux density needed to
  support the overlying corona, $^g$inferred heating rate, $^h$cooling time, 
  $^i$the wave period used
  for intermediate waves.
  \label{tab:times} 
\end{table}

\section{Calculations}

\subsection{A minimal framework}

We perform numerical calculations of a partially-ionized
hydrodynamic plasma, solving for
the evolution of mass density, momentum and energy along a tube of flux.
As is usual, we
assume that departures from thermal distribution functions are
\emph{small} because of the relatively rapid exchange of energy and
momentum between like particles \citep[e.g.][see the estimates in
Table\ref{tab:times}]{Braginskii1965}. We
also assume that all particles have the same temperature, although the
energy equilibration times between  particles of disparate mss
can exceed dynamical evolution times
(Table~\ref{tab:times}).  But unless the ion and  neutral temperatures exceed
electron temperaures by more than factors of 40 or more, only the electron
temperature controls collision times for electron-particle collisions.  
The long ionization and recombination
times for a given electron temperature can greatly exceed dynamical times, so therefore
we must solve for populations of
hydrogen and helium neutral and their ions, writing these collisional terms
as source and sink terms in the conservation equations for hydrogen and helium.
The solution ``vector'' evolved in time and space is therefore
\begin{equation} \label{eq:solvec}
  \left (\rho, \rho u, E,  n_{H}, n_{H^+}, n_{H}, n_{He}, n_{He^+}, n_{He^{++}} \right )
\end{equation}
where $\rho$ is the mass density (a sum of particle masses and number densities),
$u$ the fluid velocity, $E$ the total
energy, and $n_i$ the various number densities of the constituent ions.
The electron density $n_e$ is the sum of all
free electrons from H and He, and 
then the equation of state is used to determine the
  temperature $T$.  Being a 1D calculation, there is no explicit dynamics perpendicular to the magnetic
field. Instead, the dynamics is captured implicitly through the heating
term in the energy equation (see equations \ref{eq:volrate} and \ref{eq:massrate} below).
The intermediate  waves drive perpendicular macroscopic drifts between ions,
electrons and neutral species.  The kinetic energy of these drifts is
destroyed by random collisions leading to irreversible energy
exchange, as is readily understood as the conversion of bulk
motion (the different centroids of the distribution functions) into
heat (widths of distribution functions).
%The equivalence
%between this heating term and the more familiar approach
%of using the ``Cowling conductivity'' \citep{Cowling1956} is made explicit
%in equations (7.7) - (7.14) of \citet{Braginskii1965}, derived
%by manipulating equations of motion for electrons, ions and neutrals.

The interaction of intermediate waves with
  a partly-ionized plasma depends on various time scales, from
  collision times between abundant
  particles (predominantly atoms and ions of H and He,
  and electrons), periods of gyro-motion of ions,
  to wave oscillation periods \citep[e.g.][section 8]{Braginskii1965}.
  Table~\ref{tab:times} lists the parameters most relevant for this study,
  presented for hydrogen only, for clarity or presentation.  
  Values for time scales for helium impact with 
  other particles were used
  from \citet{Judge2007a}
  and \citet{Gilbert+Hansteen+Holzer2002}.

\tabflux

\subsection{Dissipation of intermediate waves}

Consider an intermediate 
mode with amplitude $\xi$ cm~s$^{-1}$ 
and group velocity $v_g$ propagating 
along a magnetic line of force. 
Again, for clarity we consider here a hydrogen 
plasma, below we include helium explicitly,  and treat radiation losses of trace species using a lookup table. 
We follow
section 3 of \citet{Holzer+Flaa+Leer1983}.
The flux density of energy carried by the 
wave motion is 
\begin{equation}
    F = \rho \xi^2 v_g
  \mathrm{\ \ \  erg~cm^{-2}~sec^{-1}, }
\end{equation}
For the intermediate mode, $v_g$ equals the Alfv\'en
speed $c_A = B/\sqrt{4\pi \rho_i}$. $\rho_i = n_i m_H$ is the mass
density of protons with number density $n_i$, or $c_A = B/\sqrt{4\pi
  m_H (n_i+n_n)}$ if the ions exchange momentum with ambient neutrals
with number density $n_n$, on a time scale 
which is much less than the wave period $P_W$ (Table~\ref{tab:times}).  This condition is valid
below the corona. Within the corona there are so few neutrals that one
can set $n_i \approx n_i+n_n$ so that the plasma inertia throughout
the atmosphere can be written using $m_H(n_i+n_n)$ everywhere.

The ions must be magnetized ($\Omega_i \tau_{in} \gg 1$), in order that the magnetic 
wave motions drag 
ions through neutral gas
with significant drift speeds. Here $\Omega_i$ is the ion gyro frequency \citep{Braginskii1965}:
\begin{equation}
    \Omega_i = \frac{eB}{m_i c} = 0.96\cdot 10^4 B
\end{equation}
for protons. 
In typical models with magnetic fields concentrated in kG fields in the photosphere, extending to 1-10 G in the corona, the protons are strongly magnetized across the chromosphere and corona (see Table~\ref{tab:times}).

The volumetric fluid heating rate for an ion-neutral drift speed $\overline u$ is
 \citep[see page 812 of][]{Holzer+Flaa+Leer1983}:
\begin{equation}
    \dot E_v \approx 
    \frac{1}{\tau_{in}} \rho_i {\overline u}^2. %=
    %2.5 \sqrt{T} n_n n_i m_H u_5^2 
    \mathrm{\ \ \  erg~cm^{-3}~sec^{-1}, }
\end{equation}
At low wave frequencies $\omega$ ($\omega \tau_{in} \ll 1$), collisions dominate 
the dynamics of ions and neutrals, the resulting diffusion leads to a small
drift speed $u$:
\begin{eqnarray} \label{eq:u}
  {\overline u} &\approx& \omega \tau_{in} \cdot \xi    \mathrm{\ \ \  cm~sec^{-1}, }\\
  &\ll& \xi \nonumber
\end{eqnarray}
where $\xi$ is the wave amplitude in cm~sec$^{-1}$. In the opposite
limit where $\omega \tau_{in} \gg 1$, the drift speed ${\overline u}
\approx \xi$.  Then, for both cases we can write
\begin{eqnarray}
\label{eq:volrate}
    \dot E_v &\approx &
\frac{(\omega\tau_{in})^2}{1+(\omega\tau_{in})^2} \frac{1}{\tau_{in}} \rho_i \xi^2
     \mathrm{\ \ \  erg~cm^{-3}~sec^{-1}, }
\end{eqnarray}
The heating per unit mass is 
\begin{eqnarray} \label{eq:massrate}
    \dot E_m &\approx& \frac{\dot E_v}{m_H(n_i+n_n)} \\
    &\approx& 0.9 \cdot \sqrt{T} 
     n_< \xi_5^2 
     \mathrm{\ \ \  erg~g^{-1}~sec^{-1}, }
\end{eqnarray}
in erg~g$^{-1}$sec$^{-1}$, where $n_<$ is the smaller of $n_i$ and $n_n$, and 
$\xi_5 = 10^5\xi$. $\xi_5$ is the drift speed  in \velu.  Here we have used the elastic collision rate between protons 
and H atoms, $\tau_{in}$.

Since the drift speed $u$
also constitutes an electric
current, the frictional dissipation can also be written in terms of
an electric current and a 
conductivity, which for 
motions across the field
is the perpendicular 
conductivity \citep[page 281 of][]{Braginskii1965}.
Under weakly ionized or cold plasma conditions, the latter 
can be treated using ``Cowling's
conductivity''  \citep{Cowling1956}.   The 
conductivity approach has been implemented in modern 3D MHD codes. 

Below we will study
calculations with fixed input
energy flux densities of intermediate waves which, in
the initial state, produce rms wave amplitudes of 1, 3 and 10 \velu{} within
the initially hot coronal plasma (Table \ref{tab:fluxes}), all
of which are modest 
values compared with
observations \citep[e.g.][]{2011Natur.475..477M}.
These initial amplitudes change in
time as the densities evolve during the calculations.  We made
calculations with wave periods $P_W$ of 180 and 1000 seconds
for reasons given below. 
%compatible with the calculations of \citet{2017ApJ...834...10R}.
We also make the assumption that damping is weak and use
a constant wave flux along each tube.
In section \ref{subsec:limitations} we will find that the damping lengths
\begin{equation}
L \approx c_A\tau_{in}
  \mathrm{\ \ \  cm }
\end{equation}
\citep[][eq. 16]{Holzer+Flaa+Leer1983} 
exceed the lengths of the tubes.

%\bibliographystyle{mnras}
%\bibliography{ms}
%\end{document}

 \subsection{Numerical Method }

We solve equations for the conservation of mass, momentum, energy
and the number densities listed in the vector (\ref{eq:solvec}), 
for a single fluid. The vector components are solved as a function of time and distance along a
closed flux tube.  The geometry is a simple arc of a circle which intersects
the lower atmosphere at a given height near the photosphere, with no geometric
expansion.  Three types of calculations were made, the arc center being
at the footpoint height and in a vertical plane, some calculations
with a modified gravity accounting for an arc in a plane tilted to the
vertical, and calculations where the center of the circle defining the
arc is displaced a few Mm below and above the footpoints. These
geometries serve to test how the gravity vector affects coupling
between the chromosphere ($<$ 2Mm height) and the higher regions.

All variables at each (deep) footpoint were held fixed
at the initial state, where $\log_{10}\rho\approx -9.22$ g~cm$^{-3}$ (cf.
the values much higher up in the atmosphere in Table~\ref{tab:times}). While the model
is almost symmetric around the 
mid-point, the boundaries were
set to values differing by 10\%{} in density to introduce a little asymmetry in the dynamics, since in the real Sun the two boundaries are different.
At these densities, which lie 700 km above
 the continuum photosphere of traditional models, they are about 9 pressure scale heights below the initial transition region.  
The
 conservation equations for 
for populations of H atoms, protons, helium atoms, He$^+$ and He$^{2+}$ ions include the advection and ionization and recombination terms (see  the Appendix
of  \citealp{2012ApJ...746..158J}).   
The scheme  can be seen as
a simplified version of
a multi-fluid calculation
in which the ion-neutral collisions induced 
by wave motions perpendicular to the general magnetic field direction
serve to heat the 
fluid.  

The force balance includes the pressure gradient and gravity vector projected along the direction of the flux tube, $g(s)$. 
With
the assumption of weak
dissipation,
the momentum transfer to
the fluid from the waves, 
is negligible compared 
with the gravity and pressure gradient terms
\citep[][section 3(b)] {1980ApJ...242..260H}.
The energy balance includes  the explicit 
heating terms of 
equation~(\ref{eq:volrate}),
heating due to EUV radiation from the ambient corona, and  explicit
radiation losses 
computed as a function of time for H and He, and as a lookup table for the trace species such as C, N, O etc. 
\citep{2012ApJ...746..158J}.
The energy equation is
\begin{eqnarray}
    \frac{\partial}{\partial t} E +  
    \frac{\partial}{\partial s} u\left (E +p\right)  &=&     \dot E_{V} +     \mathcal{H}_{EUV} +
    \mathcal{H}_{CHR}-\mathcal{C}_{R} 
 \\
    & &
    - \rho g(s) u + 
    \frac{\partial}{\partial s} 
    \left(
    \kappa T^{5/2}  \frac{\partial T}{\partial s}
    \right) \ \ \ \ 
\end{eqnarray}
where
\begin{equation}
     E = \sum_i \frac{n_i kT}{\gamma-1} + \frac{1}{2}\rho u^2 + U
\end{equation}
is the total energy per unit volume, the sum is over all particles, and $\gamma=3/2$ because internal energy is explicit in $U$.  $\kappa=10^{-6}$ is the Spitzer coefficient of heat conduction, 
$\mathcal{H}_{CHR}$ is a  small \textit{ad-hoc} heating term to maintain the chromosphere against strong radiation losses near the footpoints \citep{Hansteen1993}.  We used $\mathcal{H}_{CHR} = 10^{-17}$ erg~particle$^{-1}$ s$^{-1}$,
 The curves labeled ``particles'' in the following figures 
  show this term. $\mathcal{C}_{R}$ is the (positive definite) radiative loss term, and  $U$ is the latent heat of ionization computed directly from the ionization potentials and number densities of ions, including a correction 
  for the fact that hydrogen is ionized by
  the external source of radiation formed in the photosphere, in the Balmer continuum.

We use a first-order
Lax-Friedrichs explicit
integrator in time 
\citep{1996JCoPh.128...82T}.   This scheme avoids use of characteristic variables. 
The equations are written and solved in conservative form, but
 an operator splitting technique is adopted to solve for the field-aligned electron 
heat conduction
(\citealp{2012ApJ...746..158J}).  
We used uniform grids with spacings of
about 10 km, to try to resolve the transition region.   There is of course implicit dissipation owing to the numerical scheme, causing smoothing in space close to steep gradients (shocks and the steep transition region).  We found that the extension to  second-order 
scheme using total variation diminishing methods 
\citep{1996JCoPh.128...82T}, while less dissipative, was far less robust.  The conservative formulation guarantees that 
only smoothing occurs, and the final equilibrium states contain no steep 
gradients.   Our main conclusions therefore
are not strongly dependent on numerical
diffusion.

We made calculations for tube lengths $L$ of 10, 20, 30, and 40 Mm, using wave periods $P_W$
of 180 and 1000 seconds.
We used a field strength of 
$B=10$ G and 
wave flux densities $F_W$ of
$10^3, 10^4$ and $10^5$ 
erg~cm$^{-2}$~sec$^{-1}$. 
With these conditions the
largest wave amplitudes 
$\xi$ at the start of each calculation were 1, 3 and
10 \velu{} respectively.
The calculations made are summarized   Table~\ref{tab:calcs}.

\subsection{Relaxation following coronal heating}

The first calculations we made were started from a model close to force and 
energy balance with steady heating rates of $2\cdot 10^{-12}$ 
and $2 \cdot 10^{-14}$  erg~particle$^{-1}$~s$^{-1}$
for 
protons and neutrals respectively, run until 
almost static (all velocities well below all sound speeds). The goal was merely to begin from a physically realistic coronal state.
 \citet{Anderson+Athay1989} found that heating
rates $2\cdot 10^{-13}$ and  $2\cdot 10^{-14}$ 
 erg~particle$^{-1}$~s$^{-1}$ --
corresponding  to 
$10^{11}$ and $10^{10}$
 erg~g$^{-1}$~s$^{-1}$ -- 
were
required to support
the average quiet 
solar corona and chromosphere respectively.  Our initial state
(calculation ``initial''
in Table~\ref{tab:calcs})
is representative of 
a more aggressively heated 
 corona, with coronal pressures of between
 0.17 and 0.6 dyne~cm$^{-2}$. 
 In contrast, the quiet 
 Sun's coronal pressure is
 about 0.08 dyne~cm$^{-2}$. 
 The thermodynamic structure is shown in Figure~\ref{fig:start}.
  
From this initial state, the steady heating rates of protons and neutral particles that led to the formation of
 the $10^6$ K corona were set to zero, and the calculation allowed to relax,
 including only the heating the 
 through the $\dot E_V$, 
  $\mathcal{H}_{CHR}$, and for half of the cases the $\mathcal{H}_{EUV}$ terms.  In making these calculations we  attempt to model 
  the situation where intermediate waves persist, but the special conditions leading to coronal heating do not occur. 
\figinter
Figures~\ref{fig:inter} and \ref{fig:final} show intermediate and final calculations for 
a $L=20$ Mm calculation  driven by waves of $P_W=$180 second
period and wave flux $F_W$ of $10^4$ 
erg~cm$^{-2}$s$^{-1}$, which is
calculation 20 
in Table~\ref{tab:calcs}.  
The behavior seen is typical of others
listed in the Table
and marked as ``EQ'', signaling a
near-equilibrium final state.  These calculations continue to evolve very slowly, but the important
result here is that, compared with dynamical time scales, these are \textit{extremely long-lived}. 

\figfinal

During the first 
 hour of evolution
(Figure~\ref{fig:inter}), the pressure gradient supporting the initial state is reduced as the tube cools, the dynamics forming a simple
downflow to the footpoints until the upper chromosphere is reached, where a shock is formed.   This collapse towards a chromospheric state is as many other calculations have found
\citep[e.g.][]{1965ApJ...142..531F,1991ApJ...372..329C}.
But unlike earlier work, 
the interplay between  ion-neutral  heating, thermal evolution and the evolving times of recombination and ionization
reveals new classes of solutions governed by these
 non-linear interactions.  
Some of these
non-linearities are revealed by comparing calculations with and without
EUV photo-ionization
(Table \ref{tab:calcs} and Figure~\ref{fig:eq}).  For example, an increase in photo-ionization (itself a source of heating) 
decreases neutral
hydrogen densities, which in some cases leads to smaller peak temperatures 
(calculations 11,12), and 
vice-versa (calculations 1,2 or 7,8).

\figeq

Figure \ref{fig:eq} attempts to show graphically
the main results and trends listed in  Table~\ref{tab:calcs}.
All calculations shown began with the coronal initial state.   The $L=20$ Mm
parameters at the tube top are listed in the top right of each panel.  Larger
circles show larger lengths $L$, and 
the colors represent the values of $F_W$.
The difference between the EUV and no-EUV 
ionization cases are highlighted by the straight lines connecting the ``\textbf{x}-ed"
(EUV) and open circles.  The cluster of
red lines near log$T=4.7$ shows  
those calculations with large $F_W$ which move
up and to the left (near log$T=4.5$) 
when EUV radiation is included.  EUV radiation in these cases tends to increase
the final density and (perversely) the ratio of neutral to ionized hydrogen $n_H/n_p$.  The yellow
lines (log$F_W=4$) show a similar result
at lower temperatures for shorter tubes.
Yet radically different behavior is seen 
for the low $F_W =3$, $L=10$ Mm cases.  (Longer tubes underwent
a catastrophic collapse  for $F_W=3$). These two 
results highlight again the non-linear sensitivity of
the dynamical evolution to the process of
wave-driven ion-neutral collisions.  The
two different behaviors seem to indicate the presence of bifurcations towards dynamical 
attractors.

For each calculation with a  wave flux density $F_W$ in excess of $10^4$ erg~cm$^{-2}$s$^{-1}$,  a steady state was reached after a relaxation time of order $\tau_C$ ($\approx 2000$ seconds, Table~\ref{tab:times}) or longer, depending on
the tube length $L$. It should
be noted that the pressures at the cool
tube tops are always much less than
the initial starting pressures by
between 2 and 5 orders of magnitude. On the other hand these structures have lengths orders of magnitude larger than
the temperature scale height of
the classical transition region ($\approx 10-100$ km).  This difference suggests that these structures might be observable (see section \ref{subsec:observe}).
An example of  catastrophic 
cooling of a heated coronal
tube, calculation 13, is shown in Figure~\ref{fig:catas}.   This then represents another potential ``attractor''
state for these non-linear calculations.

Many of the calculations presented in
Figure~\ref{fig:eq} have ratios of
$n_H/n_p > 10^{-4}$.   Here then is
the (unanticipated) reason why the mechanism
can support long-lived, many Mm-long  tubes of cool plasma, almost 
in hydrostatic equilibrium with the 
reduced (field-aligned) solar gravity.  
The result is a surprise given the 
long-known tendency for cool loops
to collapse thermally, given ad-hoc
heating mechanisms \citep[e.g.][and the many references in the latter]{1965ApJ...142..531F,
TransitionRegion}.

In one calculation we repeated 
calculation 13 (the case of catastrophic
collapse) with a gravity half that of the Sun, i.e. for a tube inclined at 
60$^\circ$ to the vertical.  This
calculation (number 31) stabilized to 
a state somewhat similar to the equilibrium states for $L=40$ Mm.

\figcatas

Lastly, calculations with $P_W=1000$ seconds show 
essentially the same results.   The period of
these waves determines 
the  factor 
$$
\frac{(\omega\tau_{in})^2}{1+(\omega\tau_{in})^2}
$$ 
in 
equation~(\ref{eq:volrate}). 
This factor 
has little effect on the calculations, for at heights  originally occupied by the  coronal ($10^6$ K) plasma, 
this factor is one throughout the calculation.  Instead, a larger consideration concerns
the average transverse 
distance $\delta$ travelled by the ionized component of the 
fluid over $P_W$. Under collisionless conditions, 
\begin{equation}
    \delta = \frac{P_W}{2\pi} \xi 
\end{equation}
which for $\xi=10$ \velu{} is $286, 1590$ km for $P_W$ = $ 180, 1000$ respectively. 
Evidence for sinusoidal excursions of a few Mm owing to 
intermediate waves 
has been 
reported \citep{2011Natur.475..477M}.
This means that the ions potentially will
be swept over neutrals existing between neighboring magnetic
flux tubes, with different (field-aligned) thermal
histories. 

In summary, several trends are evident in the table and Figure~\ref{fig:eq}:
\begin{enumerate}
    \item The primary determinant of equilibrium or collapse is the \textit{wave energy flux density} $F_W$;
    \item The second factor 
    determining the fate of the calculation is the tube length, $L$.   No
    EQ solution was found for $L > 20$ Mm  when $F_W=10^3$ erg~cm$^{-2}$s$^{-1}$;
    \item The third factor is the presence of EUV ionizing radiation, as a source of ions, a reducer of neutrals, and as a source of heat (kinetic energy of freed electrons);
    \item EQ outcomes can occur  for long wave periods $P_W$;
 \item For a given wave flux density 
 $F_W$ both the temperatures and pressures at the tube apex are higher for longer tubes;
\item Changing the plane of the tube relative to the local vertical produces 
more stable solutions (calculation 31 shown in Figure~\ref{fig:halfg}). In this
calculation just one half of the pressure
gradient was needed to make the 
solution approach a stable equilibrium.
 \end{enumerate}

Frictional heating increases as the fluid cools, atoms slowly recombine, and ion-neutral
collisions increase, leading to an increase in heating. In many realizations, this  leads to dynamical stability.  Wave  damping lengths are found 
\textit{post-facto} to
exceed the lengths of the tubes considered 
(section \ref{subsec:limitations})
.

\fighalfg

\subsection{Evolution starting from a cool atmosphere}
\label{subsec:cool}

The solutions for tubes where heating 
sufficient to support a corona has ceased
are of interest  concerning the physical  connection of
the observed chromosphere and corona 
\citep[e.g.][]{TransitionRegion,2008ApJ...687.1388J}.  However, there is more
at stake than such a specific problem
in solar physics.  Here we address the question: 
\textit{starting from a very cool
atmosphere with very little heating, can 
this mechanism generate real temperature 
reversals in the atmosphere?}

To provide an answer, it is sufficient to demonstrate that
this can occur for just one calculation.  Ideally we would start from an
atmosphere in radiative equilibrium, but such models 
extend only to 
about  0.7 Mm above 
the continuum photosphere,
at a continuum optical depth of $10^{-5}$. 
The radiative equilibrium atmosphere is not
a good starting point from which to evolve the atmosphere of a star with convection anyway,  simply 
because acoustic waves  generated by photospheric  motions create relatively 
hotter plasma heated intermittently to above 
4000 K \citep{Carlsson+Stein1995}.

Therefore we start with the atmosphere 
obtained at the end of
calculation 3, with a maximum tube temperature of 4080 K. We ran cases with
$P_W=180$ seconds and $F_W$ $10^4$ and
$10^5$ erg~cm$^{-2}$~s$^{-1}$.  The results are listed as numbers 32 and 33
respectively.  The results are clear,
\textit{hot ($10^5$K) plasma is readily generated  by a moderate flux density of
intermediate waves, starting with plasma at temperatures below 5000K.} Given a very weakly ionized initial atmosphere (calculation 3 has $n_H/n_p > 200$),
perhaps from acoustic shock heating of the chromosphere or even 
meteoric impact, 
this mechanism is capable of generating an
 ionization fraction close to one with 
 large ($>10^7$ particles~cm$^{-3}$)
 several Mm above the chromosphere. 
Thus, \textit{the mere presence of 
intermediate waves generated by convection 
suffices to produce a high density, ionized
and therefore highly conducting atmosphere}. In this sense this very simple, inevitable mechanism
may be an important part of the process leading to stellar coronae,
in general.

\subsection{Observability}
\label{subsec:observe}

From our calculations we can readily find the brightnesses of emission
lines between two atomic levels labeled
$j$ and $i$.  Under optically thin conditions the critical quantity is the
emission coefficient integrated over frequency $\epsilon_{ j i}$:
\begin{equation}
   \epsilon_{ j i} ={\frac{h \nu_{ji}}{4\pi}}\,n_{ j}\, A_{j\to i}.
\end{equation}	 
$\epsilon_{ j i}$ is the frequency-integrated coefficient for total
isotropic emission into $4\pi$ steradians, ignoring stimulated
emission, in units of erg~cm$^{-3}$~sr$^{-1}$~s$^{-1}$,$h$ is Planck's
constant, the line center frequency is $\nu_{ji}$, $n_j$ the
population density of upper level $j$ and $A_{j\to i}$ is the Einstein
A-coefficient.  Figure \ref{fig:ec} shows emission coefficients for
the resonance lines of helium along calculation 22, chosen as a
calculation with a high pressure at the tube top, and thus it has some
of the largest computed coefficients $\epsilon_{ji}$.  The latter were
computed ignoring contributions to line emission from recombination
from He$^+$.  This assumption leads to lower limits on emission which
suffices for our purposes.
The emergent intensities are 
\begin{equation}
\label{eq:intense}
  I_\nu = \int_S ds \  \epsilon_{ji} \, \phi_\nu
\end{equation}	 
\intunu, where $ds$ is an 
integration along the LOS $S$. $\phi_\nu d\nu$ is the line profile function, specifying the emission
between frequency $\nu$ and 
$\nu+d\nu$ and normalized
to an integral over $\nu$ of 1.   To compute $\phi_\nu$ we added in quadrature 
the rms thermal and wave amplitudes $\xi$,
which, as noted above, evolve along with
the calculations.

To evaluate the emergent intensity we have a problem in that the line
of sight intersects the tube, represented by a line, at a single point
of measure zero. So, we must estimate essentially the thickness of the
tubes envisaged, which has nothing to do with the 1D calculation, but
depends on unknown properties of the original tube of plasma heated to
coronal temperatures that has suddenly received no heat.  This is
beyond current theoretical knowledge, indeed it would require us to
have solved the 70+ year-old ``coronal heating problem". This means we
must turn to observations and look for consequences which might be
refuted in future.  The highest resolution coronal images reveal some
tubes that appear to be close to a few hundred km
\citep{2020ApJ...892..134W}.  To be on the conservative side, we adopt
a path segment $ds=100$ km ($\equiv10^7$ cm).

Figures~\ref{fig:ec} and 
\ref{fig:prof}  show
emission coefficients 
and 
profiles of helium lines 
during the initial relaxation of the atmosphere from a coronal state, for calculation 22 (table~\ref{tab:calcs}).  
These profiles show the initial dynamic relaxation followed by a slow approach to a near-equilibrium state with a far lower brightness, no Doppler shifts, and broad profiles.

Typical intensities observed in the quiet Sun are 540 and 850 
\tintu{} for the 584 and 304
\AA{} lines of He and He$^+$ respectively
\citep{2004ApJ...606.1239P}.  So, 
using $ds=10^7$ cm,
we find that 
$$
10^7 \epsilon_{ji} \gta
 60,
 $$
or $\log \epsilon_{ji} \gta -5.2$ is needed 
if the mechanisms discussed here 
are to contribute 10\%{} to the observed level of line emission
from a coronal volume. 
The dashed line of Figure~\ref{fig:ec} shows this critical value.  Note that the value of $ds=100$ km 
for the integration path (equation~\ref{eq:intense}) 
corresponds to $0.1$ Mm on the plot, which is considerably wider than the narrow peaks close to $t=0$ s. During the first minute or so, 
the computed emission may indeed contribute signifcant emission during the initial dynamic phase.
This  period of line emission occurs through the work done by
gravity on coronal plasma as it collapses down to the chromosphere, compressing the
cool plasma and leading to line emission. 
However, these bright phases occur only along about 0.1-0.2 Mm of the tube length, so would appear as point sources in
current observations.  
Nevertheless, it appears that the conversion of gravitational energy to
radiation through compressive heating 
may contribute to some observable aspects of emission from
lines at transition region
temperatures.

At all times beyond a few sound crossing times (1000 seconds or longer),  the heating is dominated by
ion-neutral collisions. However,
the small 
computed emission coefficents 
at such times 
would require path lengths $ds$ between
1 and 10 Mm to approach
the mean observed intensities of the resonance lines of helium.

Figure~\ref{fig:prof} 
shows that line profiles after the
dynamic phase,
i.e. those dominated by the heating 
by intermediate waves and ion neutral collisions, become broad with widths
of tens of \velu.   
Such widths might 
conceivably contribute to broad, weak emission
seen in many profiles
of transition-region emission lines in the Sun and stars, providing perhaps 
an additional
explanation for the existence of 
such broad components
\citep[e.g.][]{1997ApJ...478..745W,2006A&A...449..759P}.

Lastly, unless integration paths greatly exceed the assumed value of 100 km, we can 
surmise that the two heating processes -- dynamic
compression driven by gravitational collapse,
and ion-neutral heating -- probably 
cannot account for 
 the much brighter and long ``cool loops'' proposed as
a solution to a long-standing problem of heating the lower
transition region
\citep{1991ApJ...372..329C,
2008ApJ...687.1388J,2014Sci...346E.315H}.

\section{Discussion}
\label{sec:discussion}

\subsection{Main results}

We have identified  
long-lived,  physically extended 
structures
with temperatures 
between 4000 and 10$^5$ K.  They are 
 characterized by having higher fractions of neutral
 particles than the 10$^6$ K corona by about 3 orders of magnitude.
They 
arise 
when previously heated 
coronal tubes of plasma and magnetic flux are no longer aggressively heated to
coronal temperatures.  These 
structures should exist when intermediate
MHD waves generated by flux emergence and convection drag ions across neutrals which have formed after recombination accompanies
plasma cooling.
Thus, these structures seem to be inevitable
in rotating, convecting stars, which appear 
to satisfy a sufficient 
condition for variable
magnetic fields to emerge from the stellar surface \citep{Spruit2011}. 

There is 
abundant observational evidence that 
heating of plasma to coronal temperatures is intermittent in time 
\citep[see for example][]{1991SoPh..133..357H,
1993ApJ...418..496K,
2012RSPTA.370.3217P}.  Further,
most theories of 
coronal heating require special conditions 
in time and space in order to dump mechanical energy as heat into the corona
\citep[e.g.][]{Parker1994}.   These special conditions (critical
twist, wave resonances, formation of current sheets, magnetic reconnection) all influence 
the magnetized plasma
in a fashion that 
the heating becomes highly time-dependent. 
Thus, if the coronal plasma suffers 
a lack of heating for a cooling time
or longer, then we must expect these 
novel structures to exist.  Further, we have found that, starting from a very cool atmosphere, reasonable conditions 
can lead to plasma in long tubes with 
temperatures exceeding
10$^5$ K.  

These new solutions  were unexpected by the author, as it is well known that 
``cool loop'' solutions are 
thermally and hence dynamically unstable (\citealp{TransitionRegion}, 
see Figure~\ref{fig:catas} for an example).   The significant ion-neutral heating that occurs 
just above the stratified  chromosphere,
along with the reduction in the field-aligned component of gravity can maintain a
long-lived, quasi-stable structure. The solutions are not 
strictly in equilibrium,
because all solutions 
have residual flows and waves resulting from
the non-linear interactions between 
heating and the evolving plasma densities.  But 
pressure gradients are almost balanced with the reduced field-aligned component of gravity

Such stable solutions  now seem 
inevitable in most convecting, 
rotating stars,
requiring only:
\begin{enumerate}
\item The presence of 
 magnetic fields emerging through the photosphere and generated by dynamo action,
    \item  The presence of perturbations of intermediate waves 
    by convection and/or global modes of oscillation, 
    propagating along these  magnetic fields, with
    \item wave amplitudes and periods compatible with solar observations,
    \item inclusion of frictional heating between protons and hydrogen atoms. 
\end{enumerate}

These conditions appear to be  \textit{sufficient} to create a minimum level
of heating in the outer atmospheres of
such stars. Thus   \textit{any star with magnetism emerging into the atmosphere, 
perturbed by convective motions, must 
possess a cool corona. } This includes low-metallicity stars, 
because the heating depends only on
the presence of hydrogen and helium.
It is in the dependence of cooling upon elements other than H and He that such stars will differ. Therefore we expect that 
these stars will have a different 
thermal structure from these solar calculations, all other things being equal, as the energy balance is one
between local heating and radiative losses from different amounts of trace species.

Returning to solar physics, 
a robust feature of these simple 1D models is ion-neutral heating that develops throughout the lengths of these flux tubes. We find that the dissipation of magnetic energy here can be  $\gta 10^8$ erg~g$^{-1}$s$^{-1}$. 
It seems possible that, especially above the solar limb, deep EUV observations might be compared with these model predictions, along the lines of recent work 
by \citet{2014ApJ...795..111H}, who found energy dissipation within hot coronal tubes occurring over the upper 80\% of the loops they observed. 
Even the absence of predicted features 
(Figure~\ref{fig:prof})
in deep
exposures would put constraints on the nature of
intermittency in coronal 
heating \citep[e.g.][]{1993ApJ...412..375L}. 
These calculations may have relevance to
stellar observations as well as solar, where 
symmetric, broad components are seen at the base of 
some transition region lines, usually explained as explosive events
\citep[e.g.][]{1997ApJ...478..745W,2000A&A...360..761P,2006A&A...449..759P}
.

\subsection{Limitations}
\label{subsec:limitations}

In adopting a single-fluid 
approximation we implicitly
assume that all particles collide  between 
different species at much faster rates than dynamical
time scales, exchanging 
momentum and energy faster 
than external forces can
separate them.  The longest time scales for collisions involve ions and neutrals. 
Thus when
$\tau_{in}$ exceeds dynamical
time scales our approximation will fail.  Crudely, taking $L \approx 10$ Mm and comparing $\tau_{in}$ with 
$\tau_S$ we find that 
$n_H > 3\cdot 10^5$ atoms~cm$^{-3}$.
Some of our EQ calculations dip below this limit (Figure~\ref{fig:eq}) but the majority 
lie above it.  
This value of $n_H$ is however a lower estimate when structures, such as shocks, exist on length scales below $1$ Mm. In  a multi-component plasma, the 
thermal force 
\citep[e.g.][]{Braginskii1965} can 
lead to a positive force 
along a temperature gradient akin to a negative heat flux under the effects of heat conduction.  While potentially important in regions of steep temperature gradients, it is likely to
be small when solutions evolve well below coronal temperatures.

Mean free paths (MFPs) of particles 
are $1/ n \sigma$, with $n$ the number density and $\sigma$ a collision cross section with similar particles.  MFPs for collisions between 
charged identical particles electrons are smaller than for gas-kinetic (neutral-neutral) MFPs. The ion-neutral and neutral-neutral MFPs are 
on the order of $10^{14}/n$
cm.   We have made calculations using  tubes of length $L \approx 20$ Mm, so that, for neutrals 
exhibiting linear motion, 
we would need 
$n_H > 10^5$ cm$^{-3}$ for the system to
be collisional.
For protons which exhibit gyro motion, we would require $n_p > 3\cdot 10^4$
cm$^{-3}$.  %\citet{Parker2007} has argued that because fluid mechanics is based upon the same set of conservation laws as are embedded in kinetic theory, fluid mechanics can nevertheless be used (with appropriate modifications of, e.g., the pressure tensor) to study
%cases where the MFP exceeds
%the system studied. 
Even in the ``collisionless''
limit, modified ``collisionless MHD equations'' can be derived. 
For example, collisionless  MHD models are applied to  Earth's magnetosphere where 
the MFP is about a light year \citep{LFM2004}.  Such modifications are not important for 
the major conclusions of the present work.

More importantly is the 
need to have many particles in each computational cell, in order that the appropriate averages are meaningful.  In our calculations we typically use a cell of at least $10$ km. In this case the averages (density, momentum, energy density) are well-defined when $n > 10^{-4}$ particles cm$^{-3}$. The end states of the catastrophic collapse
violate this condition
(Figure~\ref{fig:catas}).

The geometry is entirely
1-dimensional, taking no
account of the expansion of the tubes
of flux from the low
chromosphere and into the corona.  The expansion
leads to an upward magnetic pressure gradient within a tube.  But when $B\gta 1$G this expansion takes place almost entirely within
chromosphere, because it spans some 9 pressure scale heights.  Our goal is to study how ion-neutral collisions might lead to 
effects initially within 
the (initially) \textit{far more extended corona}.
The physics of the chromosphere is certainly 
important, setting a lower boundary condition 
for plasma evolution.  Our model simply provides a ``chromosphere'' as a  
stratified layer with 
small wave speeds. It serves as  
a reservoir of mass
which can be supplied along   flux tubes. 

Finally, the assumption of weak wave damping 
might fail.  But for all calculations made here the damping length exceeded 200 Mm in the final state. The flux of intermediate waves is almost undamped.

\subsection{Future thoughts}

This series of calculations
was in part prompted by 
remarkable data from the more extended corona, in which the He~I 1083 nm line has been 
observed to be brighter
than predicted, even with recently
updated atomic calculations 
\citep{2020arXiv200608971D}. 
With Prof. J. R. Kuhn, the author is investigating 
ion-neutral and ion-dust-neutral dynamics in the inner heliosphere, following   observational work 
obtained during total solar eclipses for more than two decades
\citep{1996ApJ...456L..67K,1998AdSpR..21..315M,2007ApJ...667L.203K,2010ApJ...722.1411M,Habbal+others2018}.  
These 
are
fundamental processes throughout astrophysics. Yet again the notoriously hot solar corona is revealing unexpected results that may be
important in astrophysics, from star formation to supernova remnants.

\section*{Acknowledgments}
\vskip 12pt

The author is grateful to Matthias Rempel for useful comments.
Prof. J. Kuhn provided many comments and encouragement in the course
of this work, without which this work would not have been completed.
The referee provided excellent insightful comments which greatly
helped the author in producing a far more readable paper.  This
material is based upon work supported by the National Center for
Atmospheric Research, which is a major facility sponsored by the
National Science Foundation under Cooperative Agreement No. 1852977.

\tabcalcs

%at: judge at ucar.edu.

\figec
\figprof

\bigskip
\noindent 
\section*{Data  availability statement}

The exploratory research reported here uses IDL-based 
software developed by the first author without documentation.  The ``data'' produced are exploratory in
nature.  Interested readers 
can request outputs from PGJ.

\bibliographystyle{mnras}
\bibliography{ms}
\end{document}